\documentclass[twocolumn,showpacs]{revtex4}
 \usepackage{amssymb} \usepackage{graphicx}

\begin{document}
 \title{Effect of Modified Dispersion Relations on Immirzi Parameter}
 \author{\"{O}. A\c{c}{\i}k}
\email{ozacik@science.ankara.edu.tr}
\author{\"{U}. Ertem }
 \email{uertem@science.ankara.edu.tr}
\address{Department of Physics,
Ankara University, Faculty of Sciences, 06100, Tando\u gan-Ankara,
Turkey\\}

\date{\today}

\begin{abstract}
Black hole entropy calculations which are based on counting of
microstates and based on modified dispersion relations in the
framework of loop quantum gravity are considered. We suggest that
the inconsistency of two approaches can be explained by different
ways. This inconsistency can affect the definition and constancy of
the Immirzi parameter or order of the modification constants of
dispersion relations. Possible results of these effects are
discussed.
\end{abstract}


\maketitle

\section{Introduction}

In Loop Quantum Gravity (LQG) \cite {Rovelli Thiemann}, there is a
free parameter $\gamma$, called Immirzi parameter that has no effect
in the classical theory but has an effect in the quantum theory. So,
it reflects the quantization ambiguity of the theory. LQG is based
on a connection formulation of general relativity whose phase space
variables are an $SU(2)$ connection and a densitized triad. However,
there is a one parameter family of canonical transformations which
lead to the same Hamiltonian formulation of general relativity. The
parameter that labels the family of canonical transformations is the
Immirzi parameter. Different values of $\gamma$ are reflected in the
different forms of the Hamiltonian constraint. If $\gamma$ is
selected as a complex number (more specifically the complex number
$i$), then the Hamiltonian constraint is simpler than in the ADM
formulation and quantization can be easier, but in this case, one
has some extra reality constraints. The pure complexity of $\gamma$
is not a necessity and if it is selected as a real number, then the
Hamiltonian constraint is more complicated, but it is still
manageable for quantization and there is no reality constraints in
this case \cite {Rovelli Thiemann, RovThi}. However, in quantum
theory $\gamma$ does not vanish and still present in the spectrum
calculations of operators. In LQG, geometric operators such as area
and volume have discrete eigenvalues and $\gamma$ appears in the
spectrum of these operators. So, finding the value of $\gamma$ means
determining the area and volume quantums. Hence, there is a
quantization ambiguity for quantum gravity and choosing a value of
the Immirzi parameter needs an explanation, namely it must be fixed
by theoretical or experimental ways.

Immirzi parameter appears also in the computation of black hole
entropy in the framework of LQG, because of the relation between the
area and entropy of a black hole. So, one can determine the value of
$\gamma$ by comparing black hole entropy-area relation found in LQG
to the semiclassical Bekenstein-Hawking (BH) entropy-area relation
\cite{ABCK}. Black hole entropy-area relation have been found by
counting of microscopic states for a fixed area value, and besides
the linear area term, there is also an $ln$ correction term in the
found relation \cite{DomLew,Meissner}.

On the other hand, in LQG, some theoretical calculations reveal the
presence of corrections to energy-momentum relations, so it implies
some modifications of dispersion relations
\cite{GamPul,MorUrr,MorUrr2,Smolin}. This kind of modification
effects may be observed at gamma ray and ultra high energy cosmic
ray threshold anomalies \cite{MagSmo}. Modified dispersion relations
can be understood in the framework of Deformed Special Relativity
(DSR) which refers that there is an observer independent invariant
energy scale, Planck energy, besides the invariant speed of light
\cite{MagSmo,AmSmSt,KowNow,KowGlik,KowNow2,FrKGSm,KGSmo}. But
modifications of dispersion relations implies some modifications to
particle localization limit and this results some corrections to
black hole entropy-area relation \cite{ACArzLM}. In this case also,
there is an $ln$ correction term, but this time a term proportional
to square root of area also exists. So, for consistency one must
consider the entropy-area relation with correction terms in
determining the Immirzi parameter $\gamma$. This indicates some
possibilities like restrictions on modifications to dispersion
relations, or different values of $\gamma$ for different scales.

In this paper, by comparing the two different approaches of finding
black hole entropy in the framework of LQG, we discuss the
explanations for the inconsistencies between the two approaches and
find some possibilities about non-constancy of Immirzi parameter and
orders of modification constants of dispersion relations.
Organization of the paper is as follows. In section 2, we summarized
the procedure of finding Immirzi parameter with counting microscopic
states of a black hole in the framework of LQG. Section 3 discusses
how the entropy-area relation can be modified with modification of
dispersion relations. In section 4, we find an equation for $\gamma$
with considering modified entropy-area relation. This section also
includes some limits to coefficients of modification terms of
dispersion relations for the consistency with fixed $\gamma$. Some
all order modifications of dispersion relations are also discussed
for comparison with coefficient limits. In section 5, possible
effects of scale dependence of $\gamma$ is argued and section 6
concludes the paper.

\section{Immirzi Parameter from Black Hole Entropy}

Black holes in general relativity obey some laws that resemble the
thermodynamical principles. In this sense, the area of the event
horizon is related to entropy. The Bekenstein-Hawking formula gives
the entropy of a black hole which is proportional to horizon area of
the black hole $A$;
\begin{eqnarray}
S=\frac{A}{4L_{p}^2}
\end{eqnarray}
where $L_{p}$ is the Planck length. A quantum theory of gravity must
provide a mechanism for microscopic states of a black hole which
explains this entropy relation. In LQG framework, the fixed horizon
area of a black hole $A$ can be obtained from different
intersections of edges of a spin network with the horizon. Spin
networks are the basis for kinematical Hilbert space of LQG and they
are eigenstates of the geometric operators. That different
possibilities of intersections constitute the microscopic states of
a black hole. Edges of a spin network are labeled by $SU(2)$
representations $j=1/2,1,3/2,...$. So, different microscopic states
represents the different intersections with different spin labels
which result the same area value. Area operator has discrete
eigenvalues and the area spectrum includes the Immirzi parameter
$\gamma$;
\begin{eqnarray}
A=8\pi\gamma L_{p}^2\sum_{i}\sqrt{j_{i}(j_{i}+1)}
\end{eqnarray}
where the sum is over intersections. Thus $\gamma$ will appear in
entropy-area relation and can be fixed by comparing with BH entropy
for large area values.

Calculations about counting of microscopic states of a black hole
has been achieved by several people \cite{DomLew,Meissner}. The
result is that entropy-area relation includes an $ln$ correction
term;
\begin{eqnarray}
S=\frac{\gamma_{0}}{\gamma}
\frac{A}{4L_{p}^2}-\frac{1}{2}\ln(\frac{A}{L_{p}^2})+O(\frac{L_{p}^2}{A})
\end{eqnarray}
where $\gamma_{0}$ satisfies the equation;
\begin{eqnarray}
\sum_{i}(2j_{i}+1) \exp(-2\pi\gamma_{0}\sqrt{j_{i}(j_{i}+1)})=1
\end{eqnarray}
The solution of this equation can be found approximately as
$\gamma_{0}=0.27398...$. By comparing with BH entropy for large
$A/L_{p}^2$ values, it can be seen that $\gamma$ must be equal to
$\gamma_{0}$. Fixing $\gamma$ means fixing the quantum of area, and
one can find from (2) that minimum possible area value of a surface.
But this determination of $\gamma$ is valid only for large area
values. If it is also valid for small area values then there can not
be correction terms rather than the $ln$ term to entropy in
calculations for finding entropy-area relations by using different
methods. But in the next section we will see that if modified
dispersion relations are considered there is a correction term to
entropy which is proportional to square root of area and this will
effect the fixing of $\gamma$.

\section{Black Hole Entropy from Modified Dispersion Relations}

Several theoretical calculations about light propagation and
neutrino propagation in LQG \cite {GamPul,MorUrr,MorUrr2,Smolin}
predict that the usual relation between energy and momentum which
comes from special relativity, may be modified at Planck scales in
the form of
\begin{eqnarray}
E^2=p^2+m^2+\alpha_1 L_{p} E^3+\alpha_2 L_{p}^2 E^4+O(L_{p}^3 E^5)
\end{eqnarray}
where $\alpha_1$ and $\alpha_2$ are constants of order one. This
kind of modification of dispersion relations can be explained by
alternative possibilities \cite{KGSmo}. Some of them are; (i) No
effect of Planck scale phenomena can be observed in low energies and
hence modification of dispersion relations has no results for
observable phenomena, (ii) Lorentz invariance breaks down and there
is a preferred frame at the Planck scale, (iii) Relativity of
inertial frames maintained but Planck length or Planck energy
becomes an observer independent quantity. This possibility is called
Deformed Special Relativity (DSR). Experimentally, modification
effects of dispersion relations may be observed by gamma ray and
ultra high energy cosmic ray thresholds \cite{MagSmo}.

Such a modification causes an effect to the Plank scale particle
localization limit \cite{ACArzLM}. An absolute limit on the
localization of a particle of energy is given by
$E\geq\frac{1}{\delta x}$. But, if one considers (5), then particle
localization limit can be found as follows;
\begin{eqnarray}
E\geq \frac{1}{\delta x}-\alpha_1 \frac{L_p}{(\delta x)^2}+
(\frac{11}{8}\alpha_{1}^2-\frac{3}{2}\alpha_2)\frac{L_{p}^2}{(\delta
x)^3}+O(\frac{L_{p}^3}{(\delta x)^4}).
\end{eqnarray}

The particle localization limit must be considered to derive the BH
entropy-area relation. So if (6) is valid then black hole entropy
relation will change because of modification terms. This has been
calculated in \cite{ACArzLM} and found that modified entropy is
\begin{eqnarray}
S\simeq\frac{A}{4L_{p}^2}+\alpha_1 \sqrt{\pi} \frac{\sqrt{A}}{L_p}
+(\frac{3}{2}\alpha_2 -\frac{11}{8}\alpha_{1}^2) \pi
\ln\frac{A}{L_{p}^2}.
\end{eqnarray}
If both $\alpha_1$ and $\alpha_2$ are vanish then entropy is equal
to BH entropy. If only $\alpha_1$ vanish then there is only the $ln$
correction term and that is consistent with the entropy corrections
which are found from counting of microscopic states (which is
mentioned in \cite{ACArzPro,ACArzLM}), but this correspondence fixes
the value of $\alpha_2$. Generally if $\alpha_1$ and $\alpha_2$ are
different from zero then there is a correction term which is
proportional to square root of area. From these discussions one can
conclude that the $\alpha_1$ coefficient of modified dispersion
relations must be zero, but we will see in the next section that
this is not the only possibility. On the other hand, the existence
of the square root area term will restrict the order of $\alpha_1$
because of the constancy of the Immirzi parameter.

\section{Immirzi Parameter from Modified Black Hole Entropy}

We have seen that there are two manifestations of black hole entropy
in the framework of LQG. One from counting of microstates and one
from modification of dispersion relations. Both includes a leading
term corresponding to BH entropy and an $ln$ correction term. But,
while there is a square root of area term in the modification of
dispersion relations approach, there is no such a term in the
counting of microstates approach. For the consistency of the two
approaches, these two entropy relations must be equal.

The second term of equation (3) and  the third term of equation (7)
are $ln$ correction terms, so we must equalize the coefficients of
these terms and we find a relation between $\alpha_1$ and $\alpha_2$
modification constants of dispersion relations
\begin{eqnarray}
12\alpha_2 - 11\alpha_{1}^2=-\frac{4}{\pi}.
\end{eqnarray}
If there is no Planck order modification to dispersion relations
namely $\alpha_1=0$, so two entropy relations are consistent, then
$\alpha_2$ must be equal to $-\frac{1}{3\pi}$. On the other hand, if
$\alpha_1 \neq 0$ then we must equalize the first term of (3) and
the first two terms of (7),
\begin{eqnarray}
\frac{\gamma_{0}}{\gamma}
\frac{A}{4L_{p}^2}=\frac{A}{4L_{p}^2}+\alpha_1 \sqrt{\pi}
\frac{\sqrt{A}}{L_p}.
\end{eqnarray}
In this case we can not fix the value of Immirzi parameter to
$\gamma_0$. By using (2) and the definition $\sum
\sqrt{j_{i}(j_{i}+1)}=J$, one can find that $\gamma$ is equal to
\begin{eqnarray}
\gamma=[\frac{1}{\sqrt{2J}\gamma_0}(\alpha_{1}\pm\sqrt{\alpha_{1}^2+2J
\gamma_0})]^{-2}.
\end{eqnarray}
So, $\gamma$ is dependent to $\alpha_1$, and it is also dependent to
$J$, but this means that the value of the Immirzi parameter changes
with the number of intersections of edges, hence with the scale
determined by the area of the corresponding surface. However, if
$J^{1/2}\gg\alpha_1$ then (10) transforms to $\gamma\simeq\gamma_0$,
namely for the large area values $\gamma$ goes to $\gamma_0$. This
is expected from the counting of microstates approach. But for the
small values of $J$, $\gamma$ is changing, this is a contradiction
with the constancy of $\gamma$. If it is constant, then it must be
equal to same quantity for all area values. The smallest value of
$J$ is comes from an edge with $j=1/2$ and it is
$J_{min}=\sqrt{3}/2$. So, if $\gamma$ is a constant and is equal to
$\gamma_0$, then $\alpha_1 \ll J_{min}^{1/2}$, and this means that
\begin{eqnarray}
\alpha_1 \ll 1.
\end{eqnarray}
So, for the constancy of $\gamma$ there is no need to $\alpha_1 =0$,
but it must be much smaller than 1.

On the other hand, if $\alpha_1$ is order one, then $\gamma$ can not
be a constant for all area values, and it changes with $J$ and
$\alpha_1$. This means that for small area values, the area spectrum
must have additional dependence on $j_i$'s and also depends on
$\alpha_1$;
\begin{eqnarray}
A=16\pi L_{p}^2 \gamma_{0}^2 (\frac{J}{\alpha_{1}\pm
\sqrt{\alpha_{1}^2+2J\gamma_0}})^2.
\end{eqnarray}
For large area values that is $J^{1/2}\gg\alpha_1$ , (12) converges
to (2) where $\gamma$ equals to $\gamma_0$. Then, in the small area
regime, the area spectrum must depends on the different
$\gamma$-sectors of the theory. Hence, if $\alpha_1$ is order one,
then $\gamma$ will be a scale-dependent parameter, and its values
are exactly determined by $\alpha_1$ and the scale of $J$.

\subsection{Comparisons with Some All-Order Dispersion Relations}

Some all-order modified dispersion relations have been considered in
the frameworks of $\kappa$-Minkowski space-time and Deformed Special
Relativity \cite{ACArzLM}. Various models predict different
$\alpha_1$ and $\alpha_2$ coefficients. For consistency, this
coefficients in the models must satisfy some requirements mentioned
above.

In the framework of $\kappa$-Minkowski space-time, dispersion
relations are given by
\begin{eqnarray}
\cosh(E/E_p)-\cosh(m/E_p)-\frac{p^2}{2E_{p}^2}\exp(E/E_p)=0\nonumber.
\end{eqnarray}
In this case $\alpha_1=-1/2$. So if this theory is true, then in the
lack of $\sqrt{A}$ term in black hole entropy from counting of
microstates, area spectrum must change with (12), namely depends on
the different $\gamma$-sectors of the theory.

Another possibility for dispersion relations is given by
\begin{eqnarray}
\cosh(\sqrt{2}E/E_p)-\cosh(\sqrt{2}m/E_p)-\frac{p^2}{E_{p}^2}\cosh(\sqrt{2}E/E_p)=0\nonumber.
\end{eqnarray}
This case has $\alpha_1=0$ and $\alpha_2=-5/18$. By vanishing of
$\alpha_1$, this is consistent with two different entropy
calculations, but the value of $\alpha_2$ is inconsistent with
predictions from consistency of entropy relations discussed above.

In the case of Deformed Special Relativity, dispersion relations are
given by
\begin{eqnarray}
\frac{E^2}{(1-E/E_p)^2}-\frac{p^2}{(1-E/E_p)^2}-m^2=0\nonumber.
\end{eqnarray}
This is the case of both $\alpha_1$ and $\alpha_2$ vanish, but still
there are some modifications to dispersion relations. Vanishing of
$\alpha_2$ is inconsistent with $ln$ correction terms in entropy
relation found from counting of microstates.

The true modification of dispersion relations can only be decided
from experiments and observations which are mentioned in
\cite{MagSmo}. Then, one can know the exact modification
coefficients and compare the results with the consistency conditions
mentioned above. On the other hand, if $\gamma$ is scale-dependent,
then this must be observed by future measurements of different scale
area values which then must have values of different
$\gamma$-sectors of the quantum theory.

\section{Possible Effects of Scale-Dependent Immirzi Parameter}

In the presence of the first order Planck scale modifications to the
dispersion relations, Immirzi parameter $\gamma$ must satisfy (10)
for the consistency of entropy relations that are calculated by two
different ways. This means that $\gamma$ depends on $J$ and has a
scale dependence. Scale dependence of $\gamma$ affects the area
spectrum and area eigenvalues must have an extra $J$ dependence. So,
for small scales area spectrum changes with different $\gamma$
values, but for $J^{1/2}\gg\alpha_1$ area eigenvalues are only
affected by multiplication with $\gamma_0$. These are also relevant
for the spectrum of volume and length operators, since they also
depend on $\gamma$, and are affected similarly by changing of
$\gamma$.

On the other hand, $\gamma$ enters the classical theory by Holst's
modification of Hilbert-Palatini action;
\begin{eqnarray}
S_H=-\frac{1}{32\pi G}\int(R_{ab}\wedge \ast
e^{ab}-\frac{\Lambda}{6}\ast 1-\frac{2}{\gamma}R_{ab}\wedge e^{ab})
\end{eqnarray}
where $\ast$ is the Hodge star operator, $e^{a}$ is 1-form basis,
$R_{ab}$ is curvature 2-form and $\Lambda$ is the cosmological
constant. The last term can be written as a sum of a torsion square
term and an exact term which is topological Nieh-Yan class
$R_{ab}\wedge e^{ab}=T^{a}\wedge T_a-d(e^{a}\wedge T_a)$
\cite{CandZan}. Second term is a boundary term, so $\gamma$ controls
the width of the fluctuations of the torsion \cite{FreidStar}. The
mean value of torsion is zero (in the non-existence of matter), but
it can fluctuate about the mean value. So, scale dependence of
$\gamma$ means scale dependence of the width of fluctuations of
torsion at the quantum level.

Coupling of spinors with (13) gives non-zero torsion and Immirzi
parameter has an effect on non-minimal fermion interaction term. The
coupling constant is dependent on $\gamma$ \cite{RovPer,FrMiTa}, but
it is shown in \cite{Mercuri} that if the inverse of the coupling
constant is equal to $\gamma$ then the last term of (13) and the
non-minimal coupling term together turn to a boundary term. So in
this case $\gamma$ has no effect on classical theory. In \cite{CTY},
it is argued that coupling (16) with quadratic spinor Lagrangian
indicates that $\gamma$ is the ratio between scalar and
pseudo-scalar contributions in the theory. With scale-dependence of
$\gamma$, this ratio also has scale dependence.

Another appearance of $\gamma$ is in Loop Quantum Cosmology (LQC)
\cite{Bojowald}. Because of the volume operator has a dependence on
$\gamma$, operator of the inverse scale factor $a^{-1}$ has also a
dependence on $\gamma$. The density operator $d=a^{-3}$ can be
constructed from the inverse scale factor \cite{Bojowald2};
\begin{eqnarray}
d_j(a)=a^{-3}p(\frac{3a^2}{\gamma L_{p}^2 j})^6
\end{eqnarray}
where $p(q)$ is a function derived in \cite{Bojowald3}. If $a^2\ll
\frac{1}{3}\gamma L_{p}^2 j$ then $d_j(a)\sim a^{12}$ and if $a^2\gg
\frac{1}{3}\gamma L_{p}^2 j$ then $d_j(a)\sim a^{-3}$. This is a
possible explanation for the inflationary phase in the early
universe without using scalar fields. But if $\gamma$ changes with
$j$ like (12) then (17) has an extra $j$ dependence and this may
effect the early evolution of the universe in the framework of LQC.

\section{Summary and Conclusion}

Immirzi parameter can be calculated from counting of microstates of
a black hole by comparing the found entropy relation with the BH
formula. In counting of microstates approach, the entropy has an
$ln$ correction term. In large scales, this correction term is
negligible and $\gamma$ is strictly equal to a number shown as
$\gamma_0$. On the other hand, black hole entropy is also calculated
from dispersion relations and modification of dispersion relations
induces some modifications to BH entropy. But, in this case one has
an additional correction term which is proportional to square root
of the area besides the $ln$ correction term. For consistency, these
two entropy relations found from different approaches must coincide.
Comparing the two entropies indicates some possibilities about the
Immirzi parameter and order of the modification constants of the
dispersion relations. These possibilities are as follows:

\begin{itemize}
  \item $\alpha_1$ must be zero, and hence no Planck order modifications
to the dispersion relations, so two entropy calculations are
consistent, but this time $\alpha_2$ must be equal to
$-\frac{1}{3\pi}$.
  \item $\alpha_1$ can be different from zero, but must be $\ll 1$, then
two approaches are consistent, and $\gamma\sim\gamma_0$.
  \item If $\alpha_1\sim1$, then the calculations for counting of microstates of a black hole
must be modified with a square root of area term.
  \item If $\alpha_1\sim1$ and counting of microstates approach is right, then
$\gamma$ must be scale-dependent and hence it has different values
for small scales and converges to $\gamma_0$ for large area values.
\end{itemize}

Each of these possibilities give rise to the consistency of the
entropy relations. The last possibility has some effects. If
$\gamma$ changes with scale, then spectrums of area and volume
operators have an extra $j$ dependence. It effects also width of
torsional fluctuations. Varying of $\gamma$ changes the spectrum of
$d_j$ operator in LQC, and effects the early evolution of the
universe. The correct case about the consistency must be decided by
the near future experiments.

\begin{acknowledgments}
This work was supported in part by the Scientific and Technical
Research Council of Turkey (T\"{U}B\.{I}TAK).
\end{acknowledgments}


\end{document}